\documentclass[preprints,review,accept,moreauthors,pdflatex]{mdpi}

\firstpage{1}
\pubvolume{xx}
\issuenum{1}
\articlenumber{5}
\usepackage{amsmath,amssymb}
\usepackage{amsmath}
\DeclareMathOperator{\arcsec}{"}
\pubyear{2019}
\copyrightyear{2019}
\history{Received: 22 June 2019; Accepted: 2 September 2019; Published: date}
\updates{yes} 

\pdfoutput=1

\Title{The~High Energy View of FR0 Radio Galaxies}

\Author{Ranieri Diego Baldi $^{1,2,}$*\orcidA{}, Eleonora Torresi $^{3}$, Giulia Migliori $^{4,5,}$\orcidB{} and~Barbara Balmaverde~$^{6}$}

\AuthorNames{Ranieri D. Baldi et~al.}

\address{%
$^{1}$ \quad Department of Physics and~Astronomy, University of Southampton, SO171BJ Highfield, UK;\\
$^{2}$ \quad Department of Physics, University of Turin, via Pietro Giuria 1, I-10125 Torino, Italy;\\
$^{3}$ \quad INAF---Astrophysics and Space Science Observatory of Bologna, Via Gobetti 101, I-40129 Bologna, Italy;\\
$^{4}$ \quad Department of Physics and Astronomy, University of Bologna, Via Gobetti 93/2, 40129 Bologna, Italy;\\
$^{5}$ \quad INAF---Institute of Radio Astronomy, Bologna, Via Gobetti 101, I-40129 Bologna, Italy;\\
$^{6}$ \quad INAF---Observatory of Turin, via Osservatorio 20, 10025 Pino Torinese, Italy\\}

\corres{Correspondence: r.baldi@soton.ac.uk}

\abstract{ A new class of low-power compact radio sources with limited jet structures, named FR~0, is~emerging from recent radio-optical surveys. This abundant population of radio galaxies, five~times more numerous than FR~Is in the~local Universe (z~$<$~0.05), represent a~potentially interesting target at~high and~very-high energies (greater~than 100~GeV), as~demonstrated by a~single case of {\it Fermi} detection. Furthermore, these radio galaxies have been recently claimed to contribute  non-negligibly to the~extra-galactic $\gamma$-ray background and~to be possible cosmic neutrino emitters.  Here, we review the~radio through X-ray properties of FR~0s to predict their high-energy emission (from MeV to~TeV), in light of the~near-future facilities operating in this band.}

\keyword{active galactic nuclei; radio galaxies; gamma-rays; jets; emission: non-thermal; GeV $\gamma$-rays;~TeV $\gamma$-rays}

\begin{document}


\section{Introduction}

Radio galaxies (RGs) associated with the~most massive black holes (BHs) in the~Universe, are important laboratories to study the~jet launching mechanism and~the~production of high and~very high energy (HE~$>$~100~MeV; VHE~$>$~100~GeV) radiation by Compton scattering due to the~presence of relativistic particles.

\begin{figure}[H]
\centering
\includegraphics[width=0.7\textwidth]{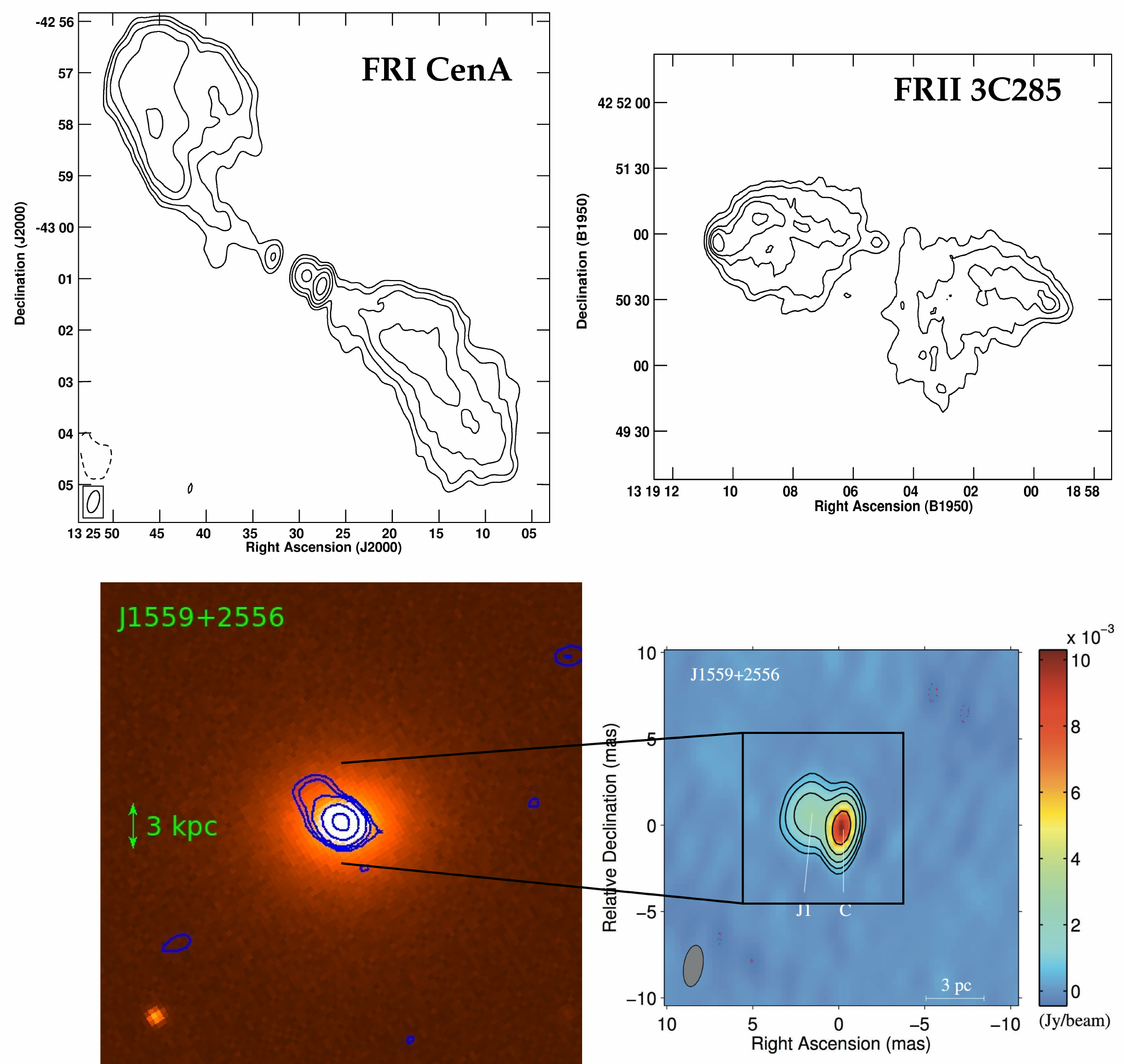}
\caption{Multi-band composite panel of radio galaxies. On the~top are two examples of radio morphologies of a~core-brightened FR~I (Centaurus~A,~\cite{burns83} at~1.4~GHz) and~an~edge-brightened FR~II (3C~285,~\cite{alexander87} at~1.4~GHz). On the~bottom, we show an~example of FR~0, J1559~+~2556. The~left panel  displays the~r-band {\it SDSS} image of the~elliptical galaxy which hosts the~FR~0 with the~blue 4.5-GHz radio contours from {\it VLA} taken from~\cite{baldi19} (3~kpc scale set by the~green~arrow). The~right panel represents the~high-resolution zoom on the~radio core (on~the~scale of~3~pc) provided by the {\it VLBI} image from~\cite{cheng18}.}
\label{fr0panel}
\end{figure}

 Although compact RGs were already known to be associated with massive early-type galaxies (ETGs) since the~early 1970s (e.g.,~\cite{rogstad69,heeschen70,ekers73,kellermann81, sadler84,wrobel91,slee94}), studies of much brighter extended RGs prevailed because of their more interesting and~richer jetted structures. These RGs are classified based on their radio morphology, distinguishing between edge-darkened (FR~I type) and~edge-brightened (FR~II type) sources~\cite{fanaroff74} (Figure~\ref{fr0panel}). Nevertheless, in the~last decade, a~renewed interest towards low-luminosity RGs (e.g.,~\cite{nyland18}) is driven by the~advent of large-area high-sensitivity optical and~radio surveys: the~cross-correlation of the~Sloan Digital Sky Survey ({\it SDSS},~\cite{abazajian09}),  the~Faint Images of the~Radio Sky at~Twenty centimetres survey ({\it FIRST},~\cite{becker95}) and~the~National Radio Astronomy Observatory Very  Large  Array  (VLA)  Sky  Survey  ({\it NVSS},~\cite{condon98}) allowed unbiased statistical  studies  of the~radio-AGN activity at~low luminosities. With this technique, the authors in~\cite{best05,best12} selected a~sample of low-power $\sim$18,000 RGs up to z~$\sim$~0.3 with radio fluxes higher than 5 mJy at~1.4~GHz, covering the~range of radio luminosity $L_{\rm 1.4 \,\,~GHz} \sim 10^{22} - 10^{26}$ W Hz$^{-1}$.  All radio morphologies are represented, including twin-jets and~core-jet FR~Is, narrow and~wide angle tails, and~FR~IIs. However, most of them ($\sim$80\%) appear unresolved or barely resolved at~the~5$\arcsec$ FIRST resolution, corresponding to a~size limit of $\sim$10~kpc.
This enormous population of low-luminosity compact RGs, different from FR~I/IIs, are still virtually unexplored.  Recently, the authors in~\cite{baldi09,baldi10}~concluded (confirmed~later by other independent radio studies,~\cite{sadler14,whittam16,miraghaei17}) that this  new population of RGs, characterized by a~paucity of extended radio emission, dominates  over FR~Is and~FR~IIs in the~local Universe. For~their radio morphological peculiarity, this class of sources has been named as~FR~0 in contrast to the~classical Fanaroff--Riley classes~\cite{ghisellini11,baldi15,baldi16}. This new class represents a~radical change in our view of the~radio sky because our past knowledge about RGs was biased towards powerful extended (from~hundreds of kpc to~Mpc) radio sources, as~selected by low-frequency high-flux surveys (see~e.g.,~the Third Cambridge (3C) catalogue~\footnote{The~3C catalogue selects radio sources with flux densities higher than~9~Jy at~178~MHz.}~\cite{spinrad85}).
However,~up~to now, the~paucity of detailed multi-band information on FR~0s limits our comprehension on the~origin of this class and~the~cause of their confined jet structures.

In this review, we summarize the~state-of-the-art of multi-wavelength studies of FR~0s,  lingering~on their emission at~high energies from keV to~TeV. In particular, we will focus on the~FR~0 population as~possible $\gamma$-ray emitters based on one case of HE detection (Tol~1326$-$379,~\cite{grandi16}), current models and~facilities, and~how their limited jets could accelerate particles that up-scatter low-energy photons or collide to eventually produce HE electromagnetic cascades.

\section{Broadband Properties: From Radio to X-Rays}

\begin{figure}[H]
\centering
\includegraphics[width=0.8\textwidth]{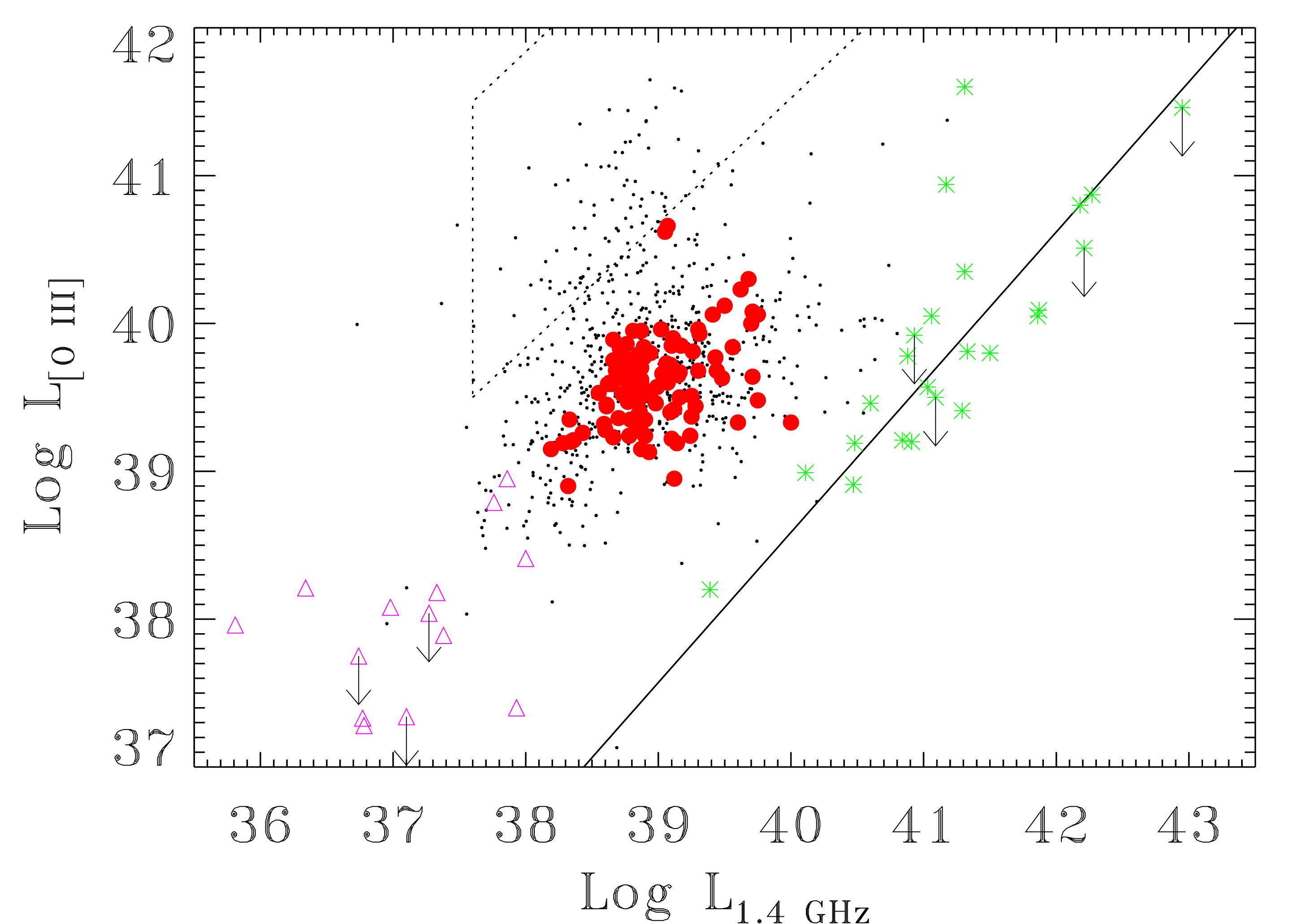}
\caption{FIRST 1.4-GHz radio vs. [O~III] line
luminosity (erg s$^{-1}$). The~small points correspond to the~low-luminosity RG sample selected by~\cite{best12}. The~solid line represents the~correlation between line and~radio-luminosity derived for the~3CR/FR~I sample (green stars). The~dotted lines include the~region where radio-quiet
Seyfert galaxies are found. The~red points are the~FR~0 included in the~FR0{\sl{CAT}}~\cite{baldi18} and~the~violet triangles are the~CoreG.}
\label{radioo3}
\end{figure}

 These compact radio sources have been carefully selected and~later included in a~catalogue, named FR0{\sl{CAT}}~\cite{baldi18}. The~selected 104 FR~0s with z~$\leq 0.05$, live in red massive ($\gtrsim 10^{11}$ M$_{\odot}$) ETGs with large BH masses ($\sim$10$^{7.5}-10^{9}$ M$_{\odot}$), and~are spectroscopically classified as~Low Excitation Galaxies (LEG)~\cite{baldi10}: these host properties are similar to those seen for the~hosts of 3C/FR~Is, but they are on average a~factor $\sim$1.6 less massive~\cite{miraghaei17,baldi18}. The~radio, optical line, X-ray luminosities of FR~0s generally match those of the~3C/FR~Is~\cite{baldi15,baldi19,torresi18a}. The~only feature distinguishing the~two classes is the~paucity of extended radio emission of FR~0s, which turns in a~core dominance~\footnote{The~core dominance is the~ratio between the~emission from the~unresolved radio core and~the~total radio emission of the~RG.} higher than classical FR~Is by a~factor $\sim$30. They show a~strong deficit of total radio emission with respect to the~3C sources, being~100~times fainter at~the~same level of (O~III) line luminosity, a~proxy of the~
bolometric AGN power~\cite{baldi10} (Figure~\ref{radioo3}). A~similar high core dominance has been also observed in nearby giant ETGs that  harbour low-power radio-loud AGN (10$^{36-38}$ erg s$^{-1}$,~\cite{baldi09,baldi16}), named Core Galaxies (CoreG,~\cite{balmaverde06a}). These~CoreG are basically miniature RGs with nuclei that are the~scaled-down version of those of FR~Is in terms of AGN bolometric luminosity, jet power and~accretion rates~\cite{balmaverde06b,balmaverde08}.

However, the~available multi-band data for FR~0s to study their nuclear activity are extremely limited. In the~radio band, generally only 1.4~GHz FIRST and~NVSS images with shallow angular resolution (5$\arcsec$ and~45$\arcsec$, respectively) were available. Hence, in the~last five years, several campaigns with the~VLA have been performed to zoom in the~GHz-band emission at~higher resolution, reaching~0.2$\arcsec$~\mbox{\cite{baldi15,baldi19}.} At~the kpc scale, the~VLA analysis reveals that FR~0s remain compact below $\lesssim$1~kpc, with only $\sim$25\% of the~sample able to emanate twin or one-sided jets extended for at~most a~few kpc (Figure~\ref{fr0panel} as~an~example). At the parsec scale, higher-resolution radio observations with the~Very Large Baseline Interferometer ({\it VLBI})~\cite{cheng18} show resolved radio jets for~80\% of the~sample. The~VLBI multi-epoch data and~the~symmetry of the~radio structures indicate that the~jet bulk speed is mildly relativistic (between 0.23 $c$ and~0.49 $c$)  with Doppler-boosting factors ranging from~1.7 to~6.

Including low-frequency radio data-points from the~Low-Frequency Array ({\it LOFAR}) and~the Giant Metrewave Radio Telescope ({\it GMRT}) surveys, the~FR~0 radio spectra from hundreds of~MHz to several~GHz are typically flat (\cite{baldi15,baldi19,capettigmrt}). However, their SEDs can be generally described as~a~gentle spectral curvature towards higher radio frequencies, broader than what is seen in young radio sources. The~flat-spectrum radio core powers tightly correlate with the~AGN bolometric luminosities, following the~relation of FR~Is and~CoreG~\cite{baldi15,baldi19}: this clearly indicates a~common central engine, where the~jet is the~main contributor to the~whole bolometric power. At higher radio frequencies, a survey at 15 GHz has been planned with the Arcminute Microkelvin Imager ({\it AMI}) array.

In the~optical band, the~continuum and~spectral information of the~nuclei are limited to the~SDSS data, which yield to a~low luminosity of the~central source and~a~prominent radio loudness~\footnote{Radio loudness, $R$, is defined  as~ratio between the~flux densities in the~radio  (6~cm, 5~GHz) and~in the~optical band (4400~\AA)~\cite{kellermann89}; where radio-loud AGN have $R~>$ 10. Assuming the~optical galaxy emission as~an upper limit on the~optical nuclear component and~5~mJy as~minimum radio flux, the~radio loudness of FR~0s is at~least $>$11.}.

\subsection{X-Ray Properties of FR~0s \label{sec2.1}}

\begin{figure}[H]
\centering
\includegraphics[width=0.6\textwidth,angle=0]{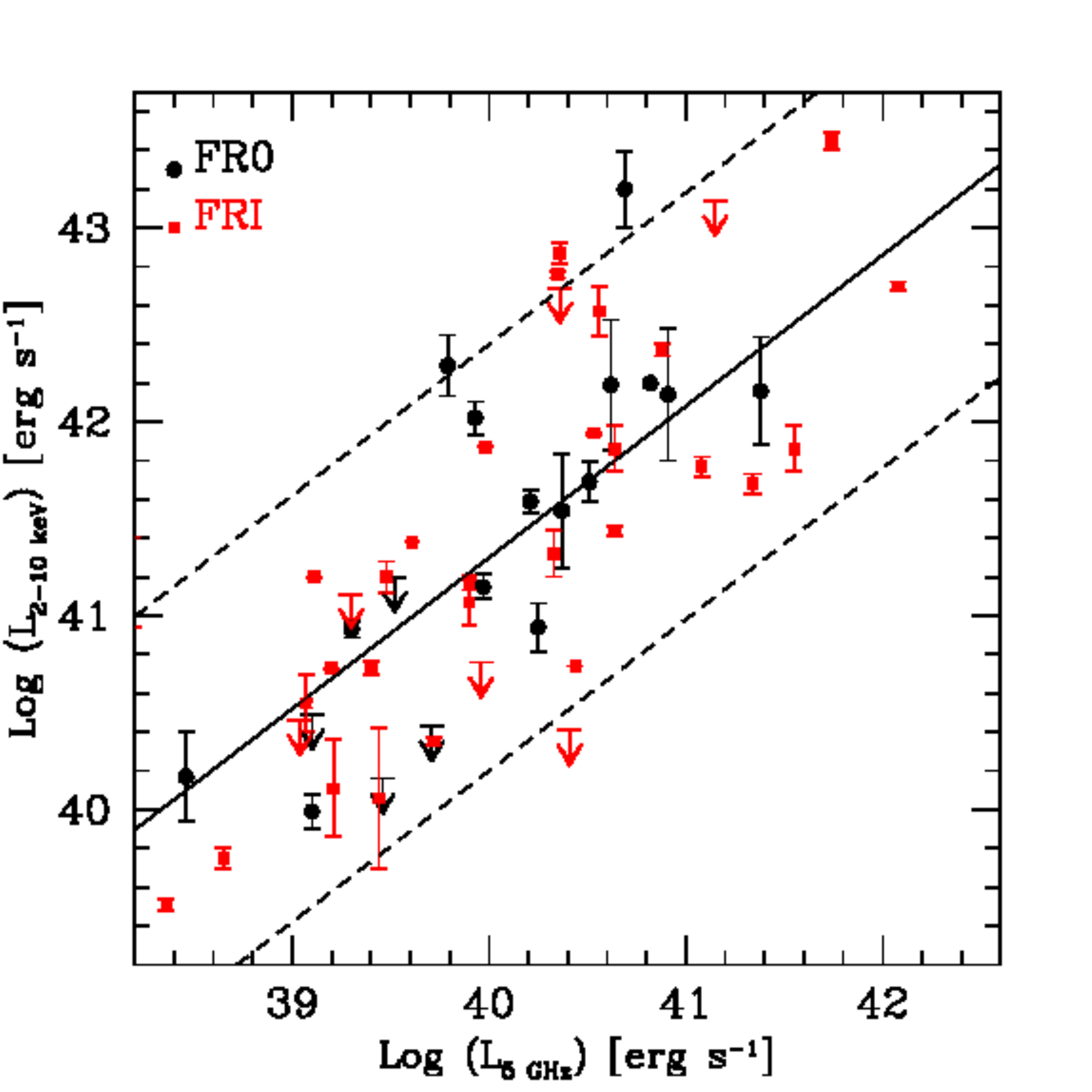}
\caption{X-ray (2--10~keV) luminosity versus radio core (5~GHz) luminosity (in erg s$^{-1}$) for FR~0s (black circles) and~FR~Is (red squares). Upper limits are represented by arrows. The~black solid line is the~linear regression for FR~0s$~+~$FR~Is: log $L_{\rm X}$~=~(7.8~$\pm$~0.6)~+~(0.8~$\pm$~0.1)log L$_{\rm 5~GHz}$. Black dashed lines are uncertainties on the~slope, taken from~\cite{torresi18a}.}
\label{loiiilx}
\end{figure}

At higher energies, the~properties of FR~0s are almost unexplored.~The authors in \cite{torresi18a} performed the~first systematic study in the~X-ray (2--10 keV) band of a~sample of 19 FR~0 galaxies selected from~\cite{best12} on the~basis of: (i)~redshift z~$\leq$~0.15, (ii)~radio size $\leq$10~kpc (in {\it FIRST} images); (iii) {\it FIRST}  flux $>$30~mJy; (iv)~LEG~optical classification; (v)~having available X-ray data in the~public archives of the~{\it XMM-Newton, Chandra} and~{\it Swift} satellites.  The~aforementioned criteria guaranteed that the~considered objects are classified as~FR~0s, although at~higher radio flux densities than the~FR0{\sl{CAT}} sources. Finally,~Tol~1326-379, the~first FR~0 radio galaxy detected in $\gamma$-rays (see~Section~\ref{sec4}), was added to the~sample. The~principal results of this study are:
\begin{enumerate}[leftmargin=*,labelsep=5mm]
    \item[1.] The~X-ray spectra of FR~0s are generally well represented by a~power-law absorbed by a Galactic column density. An additional intrinsic absorber is not required by the~data suggesting that, in these sources, the~circum-nuclear environment is depleted of cold matter (e.g.,~the dusty torus is missing).
    In some cases, the~addition of a~thermal component is required by the~data: this soft X-ray emission could be related to extended intergalactic medium or to the~hot corona typical of ETGs~\cite{fabbiano92}. The~spectral slope of the~power-law is generally steep, $<\Gamma>$~=~1.9~$\pm$~0.3. Only in two cases, which includes Tol~1326$-$379, 
    is the~photon index flatter, $\sim$1.2;
    \item[2.] FR~0s span a~range in X-ray luminosity L$_{X}$ = 10$^{40}$--10$^{43}$~erg~s$^{-1}$, similar to FR~Is.
    When the~X-ray luminosity is compared to the~radio core one, a~clear correlation is attested. This points towards a~non-thermal origin (i.e.,~the~jet) of the~X-ray emission in FR~0s as~commonly believed in FR~Is (e.g.,~\cite{balmaverde06b,hardcastle00,hardcastle09}) (Figure~\ref{loiiilx});
    \item[3.] the~central engine of FR~0s is probably powered by radiatively-inefficient accretion disc (i.e.,~Advection Dominated Accretion Flow [ADAF] model,~\cite{narayan95}), as~suggested by the~small values of the~Eddington-scaled luminosities, $\dot{L}$ = L$_{bol}$/L$_{Edd}$$\sim$10$^{-3}$\--$10^{-5}$.
\end{enumerate}

\subsection{The Uniqueness of the~FR~0 Class}

What makes FR~0 class important is their large abundance, being five times more numerous than 3C/FR~Is at~z $<$0.05. Even at~low radio frequencies where optically-thin radio jet emission is expected to dominate over the~flat-spectrum core emission, in the~LOFAR survey (the Low-Frequency Array Two-Metre Sky Survey {\it LoTSS},~\cite{shimwell19}), $\sim$70\% of RGs appears to be low-luminosity unresolved sources~\cite{hardcastle19,sabater19} at~the~same angular resolution of FIRST images (5$\arcsec$): about a~few thousand of these compact RGs are potentially FR~0 candidates with respect to a~comparable number of extended FR~Is and~FR~IIs~\cite{mingo19}.

 What makes this class of RGs unique is their lack of substantial extended radio emission and~consequent high core dominance although they harbour an~FR~I-like nucleus. {\it Are they low-power blazars?}  No, the~more-preferentially symmetric jet structures of the~FR~0s on a kpc scale rule out a significant projection effect~\cite{baldi15,baldi19}.  The~similar radio core-bolometric luminosity correlations, derived from [O~III], X-ray, of FR~0s and~FR~Is yield to their common radio misalignment, since emission line and~HE emission are typically considered independent of orientation. Another proof of the~misaligned nature of FR~0s comes from the~distribution of the~ratio between [O~III] and~X-ray luminosities (R$_{[OIII]/X}$). While FR~0s and~FR~Is show similar values of R$_{[OIII]/X}\sim$ $-$1.7 (e.g.,~\cite{buttiglione10,hardcastle09}), low-luminosity blazars have R$_{[OIII]/X}\sim$ $-$3.3 (e.g.,~\cite{capetti15,torresi18a}) as~expected if the~X-ray emission is beamed.  {\it Are they small fading FR~Is?} A past FR~I radio activity is ruled out by the~non-detection of extended diffuse emission at~low radio frequencies and~low angular resolutions ({\it LOFAR} and~{\it GMRT} \cite{capettigmrt}. The~FR~0 nuclear affinity with FR~Is is, furthermore, downsized by the~the mildly relativistic jet speed measured from VLBI with respect to the~still relativistic speeds at~parsec scales of FR~Is~\cite{venturi95,giovannini01,kharb12}. {\it Are they young FR~Is?} An age scenario, for which FR~0s are young RGs that will all eventually evolve into extended radio sources cannot be generally reconciled with the~larger space density of FR~0s than FR~Is~\cite{baldi19}. Nevertheless, a~possible contribution of $\sim$10\% of genuinely young radio sources cannot be excluded among FR~0s (see~\cite{sadler14}), based on the~fraction of sources with inverted radio spectra \cite{capettigmrt}. Unlike FR~0s, genuinely young radio sources show different multi-band characteristics:  (i)~GHz-peaked radio spectra due to strong synchrotron self-absorption~\cite{odea91}; (ii)~typically brighter in radio and~optical line due to the~enhanced emission caused by an~expanding radio cocoon in a~dense gas-rich environment~\cite{odea98}; (iii)~FR~II-like bolometric and~HE powers associated with a~radiatively efficient accretion disc (e.g.,~\cite{guainazzi06,vink06,labiano08,siemiginowska08,tengstrand09,migliori14}); (iv)~often show signatures of intrinsic H~I and~X-ray absorption~\cite{vink06,morganti18}, which suggest that these sources are typically embedded in a~rich interstellar medium.

All these characteristics and~differences with respect to the~other FR classes, blazars, young sources and~radio-quiet AGNs point to FR0s as~a~stand-alone class, which does not fully fit in a~orientation-dependent AGN unification scheme~\cite{urry95} a~priori, but does in a possible evolutionary AGN scheme \cite{garofalo19}. Furthermore, no clear incongruity in the~nuclear and~host properties of the~FR~0s with respect to the~other RGs can account for the~different radio properties.

A similar class of low-luminosity compact RGs (LLC) have been recently addressed by~\cite{kunert09}, as~kpc-scale radio sources with possibly fading steep-spectrum disrupted jets and~short-lived activity because of a~lack of significant fuelling onto the~BH~\cite{kunert10,kunert14}. These sources appear more consistent with an~interpretation of FR~I/FR~II short-lived progenitors, rather than an~intrinsically different radio class as~it emerges from the~multi-band studies of LCC. Nevertheless, a~large overlap between the~FR~0 and~LLC populations  cannot be excluded.

\section{What Causes the~Deficit of Extended Radio Emission in FR~0s?}

While {\it VLBI} observations~\cite{cheng18} and~forthcoming enhanced Multi-Element Radio
Linked Interferometer Network ({\it eMERLIN}) \cite{baldiemerlin}
observations show a~larger fraction of asymmetric pc-scale radio structures, in JVLA maps~\cite{baldi15,baldi19}, the~jets appear more symmetric at~larger scales. In addition, the~nuclear luminosities of FR~0s reconcile with those of FR~Is, suggesting similar AGN energetics. Hence, these results concur on the~picture that FR~0 jets are probably launched relativistic at~pc scale, while, at~a larger scale, they decelerate showing smaller Lorentz factors $\Gamma_{jet}$~\footnote{The~Lorentz factor for jets is \( \Gamma_{jet}~=~\frac{1}{\sqrt{1 - (v_{jet}/c)^2}} \) where v$_{jet}$ is the~jet speed.}.

Assuming a~stationary jet scenario, the~idea of a~jet stratification with different velocity patterns~\cite{ghisellini05} seems to better reproduce the~general properties of the~FR~0 class. The~similarity of the~nuclear properties of FR~Is and~FR~0s suggests that, within a~few parsecs, the~jet physics of the~two FR classes need to be analogous. Therefore, as~valid for FR~Is, we speculate that a~inner relativistic spine is likely to be responsible to produce the~multi-band properties of the~unresolved radio core and~observed pc-scale jet asymmetry. However, to account for the~confined jet capabilities of the~FR~0s, the~inner spine should be limited in length along the~jet ($\lesssim$1 kpc). The~outer layer travelling with a~mild relativistic speed ($\sim$0.3 $c$) will dominate on a kpc scale and~will account for the~observed extended jet morphologies. The~mild relativistic jet bulk speed and~a~possible intrinsic spine weakness together with entrainment of the~galaxy medium could concur to a~possible loss of jet stability. In this scenario, the~jets would suffer from deceleration and~premature disruption before escaping the~host core radius, slowly burrowing their way into the~external medium.  Nevertheless, the~optical host magnitudes of the~FR~0s, a~proxy for local density, comparable  with those of FR~Is, do not concur with the~idea of a~dense galaxy-scale environment which could cause the~jet frustration through the~interaction with interstellar density~\cite{ledlow96,kaiser07}.  Alternatively, the~scarcity of sites of plasma accelerations along the~jet could also account for their restrained jet capabilities.

 In a~time-dependent jet scenario, an~intermittent nuclear activity injects sporadically relativistic plasma in the~jet and~hence does not provide sufficient bulk flow to sustain the~jet along large distances from the~BH. This lack of plasma injection might reduce the~jet momentum and~cause significant jet deceleration within the~galaxy. To infer the~duty cycles of FR~0s by comparing the~space densities of FR~0s with respect to the~extended FR~Is, short active phases of a~few thousands years are strongly favoured with respect to longer ones~\cite{baldi18}. Analogously to FR~0s, LLC have also been interpreted as~powered by a~short-lived  outburst of the~central activity, which may turn in compact radio jets~\cite{kunert10}. Deep Low-frequency radio observations with {\it LOFAR} could reveal past radio activities of FR~0s and~measure their activity recurrence. Nevertheless,~a~preliminary analysis on a~small number of sources in the~LoTSS survey seems to reject this hypothesis.

A further scenario proposes that FR~0s have low, prograde BH spin, which would limit the~extracting energy to launch the~jet and~eventually evolve into full-fledged FR~Is thanks to the~increasing angular momentum of the~accreting material~\cite{garofalo19}.  Assuming a~BH spin-$\Gamma_{jet}$ dependence (e.g.,~\cite{tchekhovskoy10,maraschi12}), the~low BH spin would be the~ultimate reason for the~lower bulk Lorentz factor, reducing~the~ability of the~FR~0s' jets to penetrate the~galaxy medium. The~low BH spin could be caused by a~possibly poorer large-scale environment than that of FR~Is \cite{capettienvir}, which sets (i)~a~less intense accretion history than FR~Is, or (ii)~a~less frequent merger  history~\cite{dubois14} to favor BH--BH coalescence, which spins up the~central BH, a~condition required to produce extended relativistic~jets.

\section{High Energy (Gev-Tev) View of FR~0s \label{sec4}}

While the~previous $\gamma$-ray mission, {\it EGRET}, detected a~few RGs (e.g.,~\cite{steinle98,mukherjee02}), the~{\it Fermi} $\gamma$-ray space telescope~\cite{atwood09} has revolutionised our knowledge of the~sky above 100~MeV. Among the~outstanding results reached in ten years of survey, there is certainly the~discovery of RGs as~an~whole class of~GeV emitters~\cite{abdo10a}. Contrarily to blazars, where relativistic effects boost and~blue-shift the~emission of the~jet that is closely aligned to the~observer's line of sight, in RGs, where the~viewing angle of the~jet is large ($>$1/$\Gamma_{jet}$), the~detection of such sources at~high energies is disfavoured. However, misaligned RGs were also detected by {\it Fermi}-LAT in only 15 months of survey~\cite{abdo10a} and~their number continues to increase~\cite{sahakyan18,fermi19}. A few of them are also known as~VHE emitters including M87 (d~$\sim$~16 Mpc), the~first extra-galactic source detected at~VHE energies, and~Centaurus~A (CenA), the~nearest (d~$\sim$~4 Mpc) AGN to us~\footnote{In addition to M~87 and~CenA, there are other four radio galaxies detected at~TeV energies, i.e.,~NGC~1275 (d~$\sim$~76.7~Mpc), IC~310 (d~$\sim$~82.8~Mpc), PKS~0625-35 (d~$\sim$~245~Mpc) and~3C~264 (d~$\sim$~95.1~Mpc).}. Although misaligned RGs represent 2\% of the~entire extra-galactic $\gamma$-ray population that is dominated by blazars~\cite{ackernmann15,acero15}, they are extremely relevant in addressing problems related to the~jet structure, the~particle acceleration mechanism and~the~location of the~acceleration sites. Moreover, RGs have been known to significantly contribute to the~extra-galactic diffuse $\gamma$-ray background~\cite{stawarz06,dimauro18} and,~interestingly, they are often invoked as~possible sources of high-energy neutrinos and~ultra-high energy cosmic-rays~\cite{manheim95,eichmann18}.

\begin{figure}[H]
\centering
\begin{tabular}{cc}
\includegraphics[width=0.48\textwidth]{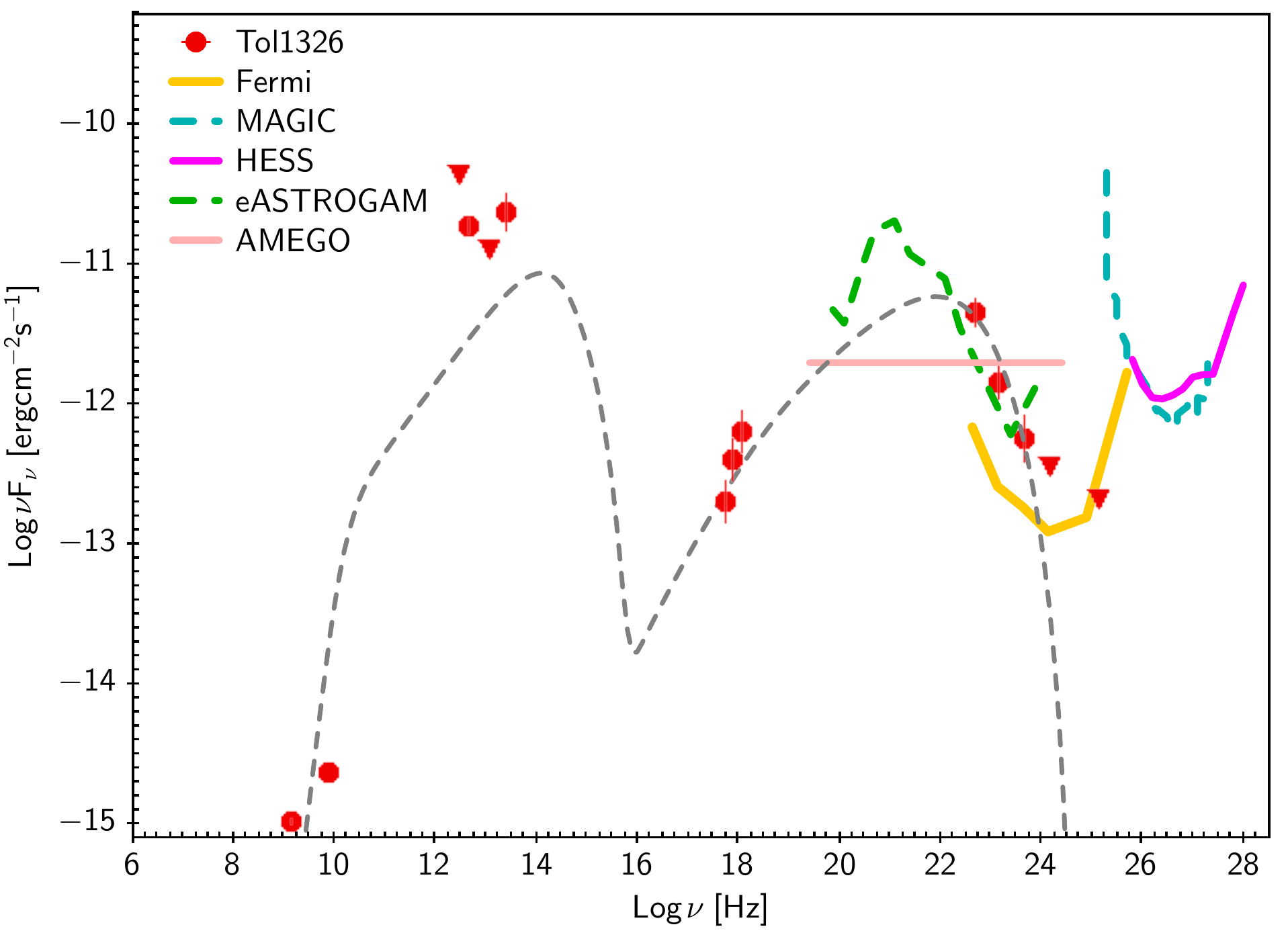} &
\includegraphics[width=0.48\textwidth]{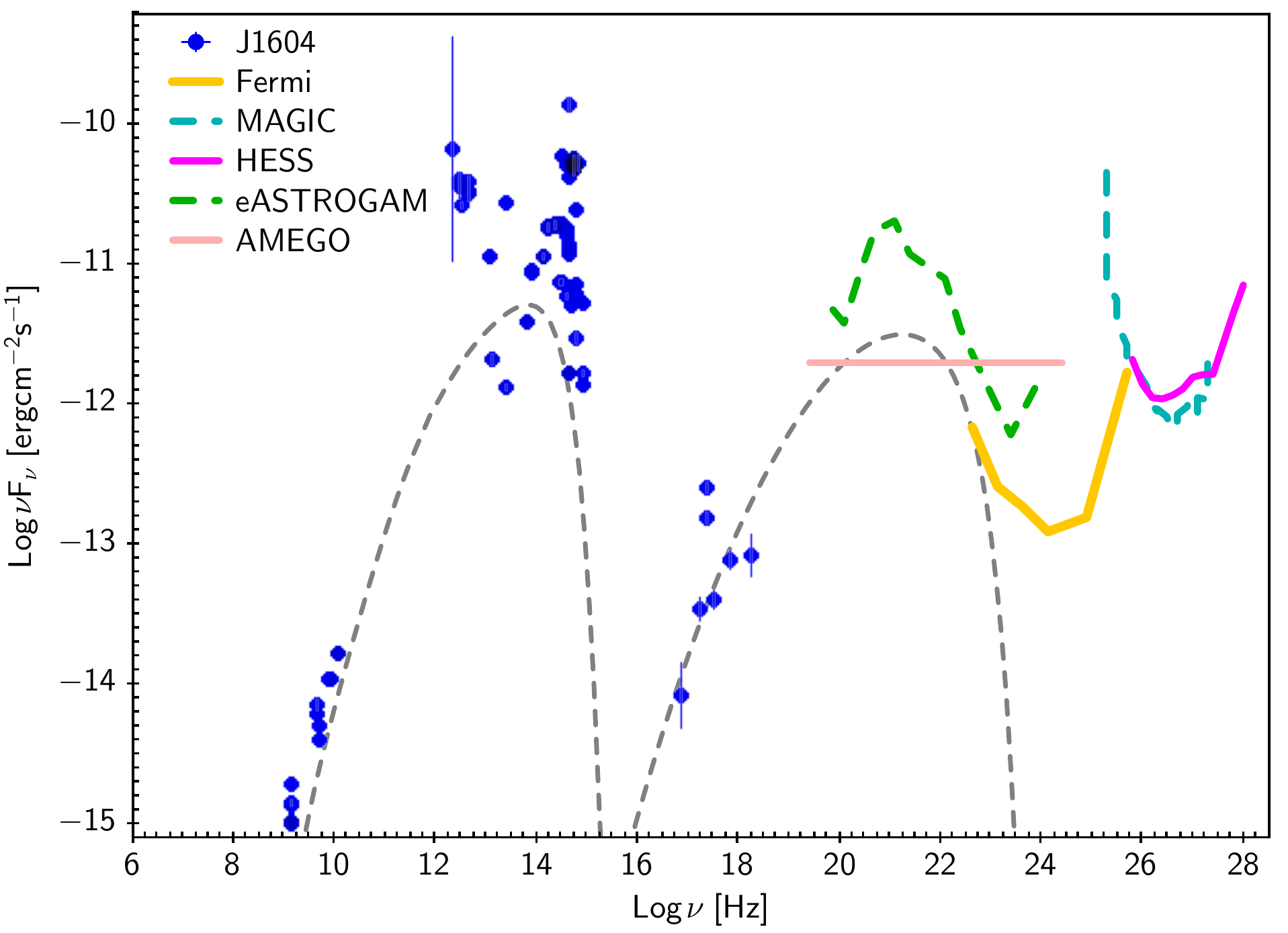}\\
(\textbf{a}) & (\textbf{b})\\
\end{tabular}
\caption{(\textbf{a}) the~SED of Tol~1326~$-$~379 (red points), the~only FR~0 observed by {\it Fermi} so far, is modelled with a~one-zone synchrotron--SSC model (grey dashed line). Following the~modelling presented in~\cite{tavecchio18}, a~moderate viewing angle ($\sim$30$^{\circ}$) and~bulk motion ($\Gamma_{jet} \sim$~2) have been assumed. (\textbf{b}) the~SED of another FR~0, J160426.51~$+$~17 (blue points) with a~model similar to that applied to Tol~1326~$-$~379 is shown (data from~\cite{grandi16}). The~expected sensitivity curves of current and~future instruments in the~MeV to~TeV window are plotted (see~the~legend). While, in the~framework of standard, one-zone leptonic models, a~detection in the~VHE range seems unlikely, both sources could be detected at~MeV energies by future missions such as~{\it AMEGO}~\cite{moiseev17} and~{\it e-ASTROGAM}~\cite{deangelis17} (see~Section~\ref{sec4.1}).
}
\label{FR0SED}
\end{figure}

Being an~emerging class of low-luminosity RGs, so far, only a few studies have focused on FR~0s in the~GeV domain. Currently, only one FR~0 has been associated with a~$\gamma$-ray counterpart by {\it Fermi}, i.e.,~Tol~1326$-$379~\cite{grandi16}. The~source is characterized by a~GeV luminosity (L$_{>1~GeV}$~=~2~$\times$~10$^{42}$~erg~s$^{-1}$) lower than blazars, consistent with  FR~Is but with a~slightly steeper photon index ($\Gamma_{\gamma}$~=~2.8) than classical FR~Is.
The~radio-GeV spectral energy distribution (SED) of the~core is double-humped (see~Figure~\ref{FR0SED}a) 
 similar to other jet-dominated radio-loud AGN. Despite the~similar radio luminosity, M~87 is less luminous than Tol~1326$-$379 by a~factor of~30 at~1~GeV and~has a~flatter SED in the~$\gamma$-ray domain. In contrast, the~SEDs of CenA and~Tol~1326$-$379 are quite similar in shape with a~steep $\gamma$-ray trend, but the~former source is about two orders of magnitude fainter.

At low energies, the~non-thermal synchrotron component of Tol~1326$-$379 dominates over any accretion-related contribution, further supporting the~case of an~ADAF disc suggested by the~X-ray study (see~\cite{torresi18a} and~Section~\ref{sec2.1}). The~peak at~the~high-energies (so-called Compton peak) appears to be relatively prominent, comparable with or even higher than the~synchrotron one, differently from BL Lac sources. The~whole SED can be consistently reproduced by a~synchrotron and~synchrotron-self-Compton (SSC) model assuming either an~aligned  ($\theta_{v} < 1/\Gamma_{jet}$) or a~misaligned ($\theta_{v}\sim$~30$^{\circ}$,~\cite{tavecchio18}) jet, with~a~total jet energy flux of the~order of few 10$^{44}$ erg s$^{-1}$~\cite{tavecchio18}.

{\it Is Tol~1326$-$379 a~unique $\gamma$-ray case?} It has been estimated that core-dominated RGs nearby, i.e.,~FR0s and~Core Galaxies~\cite{balmaverde06a}, can account for $\sim$4--18\% of the~unresolved $\gamma$-ray background below 50~GeV observed by the~LAT instrument on-board {\it Fermi}~\cite{stecker19}. Unfortunately, no FR~0 is listed among the~$\sim$37 non-blazar AGN in the~recently released Fourth LAT AGN Catalog (4LAC,~\cite{fermi19}). Nevertheless, one should note that Tol~1326$-$379 was initially classified as~a~blazar and~a~careful analysis of its multi-wavelength properties revealed its FR~0 nature~\cite{grandi16}. Furthermore, the~4LAC includes a~large number ($\sim$1090) of sources defined as~blazar candidates of uncertain type (BCUs), which~could hide new $\gamma$-ray FR~0s.  The~range of $\gamma$-ray fluxes and~photon indexes of BCUs ($\Gamma_{\gamma}$~=~1.5--3.0) are also compatible with those of Tol~1326$-$379, which can be taken as~a~reference for FR~0s at~high energies. The~cross-correlation of the~BCUs from the~4LAC with the~FR0{\sl{CAT}} did not return any match. However, the~FR0{\sl{CAT}} selected only sources in the~northern hemisphere, while more than 50\% of the~BCUs are in the~less explored southern sky, where Tol~1326$-$379 itself has been found. Thus, future efforts must be devoted to extend the~search for $\gamma$-ray FR~0 sources to this part of the~sky (see~\cite{sadler14}).

Interestingly, spectral fitting of the~synchrotron peak of the~SED of the~BCUs in the~4LAC reveals an~emerging population of faint, BL~Lac-like sources characterized by a~low synchrotron peak ($10^{13\text{--}14}$~Hz,~\cite{fermi19}). A~fraction of these sources could be truly faint low-power BL Lacs (rather~than located at~high~redshift) and~possibly represent the~aligned counterparts of FR~0s. The~joint study of the~two classes can be an~effective way to explore the~jet formation and~the~emission processes at~the~lowest levels of accretion onto the~supermassive BH (see~\cite{capetti15}).

Finally, it is important to stress that the~{\it Fermi} catalogues place a~5$\sigma$ threshold to the~detection of a~source, which might leave out inherently faint $\gamma$-ray emitters (see~\cite{migliori16} for an~example). Therefore, progress in the~characterization of the~FR~0 class in the~$\gamma$-ray window requires ad hoc analysis of the~individual targets, exploiting the~whole {\it Fermi} dataset, the~most up-to-date calibrations and~data selection, to look for sub-threshold emission. Besides the~likelihood analysis of individual sources, the~$\gamma$-ray properties of FR~0s can also be studied collectively by means of techniques searching for a~signal in excess over the~background in stacked data of a~large sample (see~\cite{ackermann16} as~example). These~procedures could unearth a~non-negligible population of low-luminosity RGs and~FR~0s emitting at~HE~\cite{best19}.

\subsection{Perspectives with Upcoming MeV--TeV Observatories \label{sec4.1}}

Here, we test whether FR~0s could be detected by current and~future generations of HE and~VHE telescopes. We use two FR~0s as~reference (Figure~\ref{FR0SED}): while one is Tol~1326~$-$~379, detected by {\it Fermi}, the~other source is a~nearby (z=0.04) FR~0 radio galaxy, J160426.51$+$17. It shows a~flat X-ray spectral index ($\Gamma_X~=~$1.1~$\pm$~0.3) and~X-ray luminosity similar to Tol~1326~$-$~379 (see~\cite{torresi18a}) that make it a~potential candidate for a~$\gamma$-ray detection. As an~exercise, we adjusted the~model adopted for Tol~1326~$-$~379 to reproduce the~SED of J160426.51$+$17 (Figure~\ref{FR0SED}b): the~output in the~GeV window could be just at~the~{\it Fermi} detection~limit.

Of equal importance for the~characterization of the~HE output of these sources will be the~synergies between the~VHE and~the~rest of the~HE window. In particular, the~{\it Fermi} observations have shown that, in RGs, the~0.1--100~GeV band probes the~Compton component beyond the~peak, i.e.,~the~decaying tail, while the~peak likely falls in the~MeV energy range. Therefore, missions such as~{\it AMEGO}~\cite{moiseev17} and~{\it e-ASTROGAM}~\cite{deangelis17}, proposed with the~goal of exploring the~MeV sky, will be crucial to make progress in our knowledge of the~processes producing the~most energetic radiation.  As~an~example, in Figure~\ref{FR0SED}, we~plotted the~expected sensitivity curves for the~two instruments on the~SEDs of Tol~1326~$-$~379 and~J160426.51$~+~$17. In~both cases, the~two telescopes would detect the~two FR~0s at~MeV energies. By~filling the~gap between the~X-ray and~GeV--TeV band, MeV measurements will also play an~important role to discriminate between leptonic and~hadronic models (see,~e.g.,~\cite{petro15}).

\begin{figure}[H]
\centering
\includegraphics[width=0.6\textwidth,angle=0]{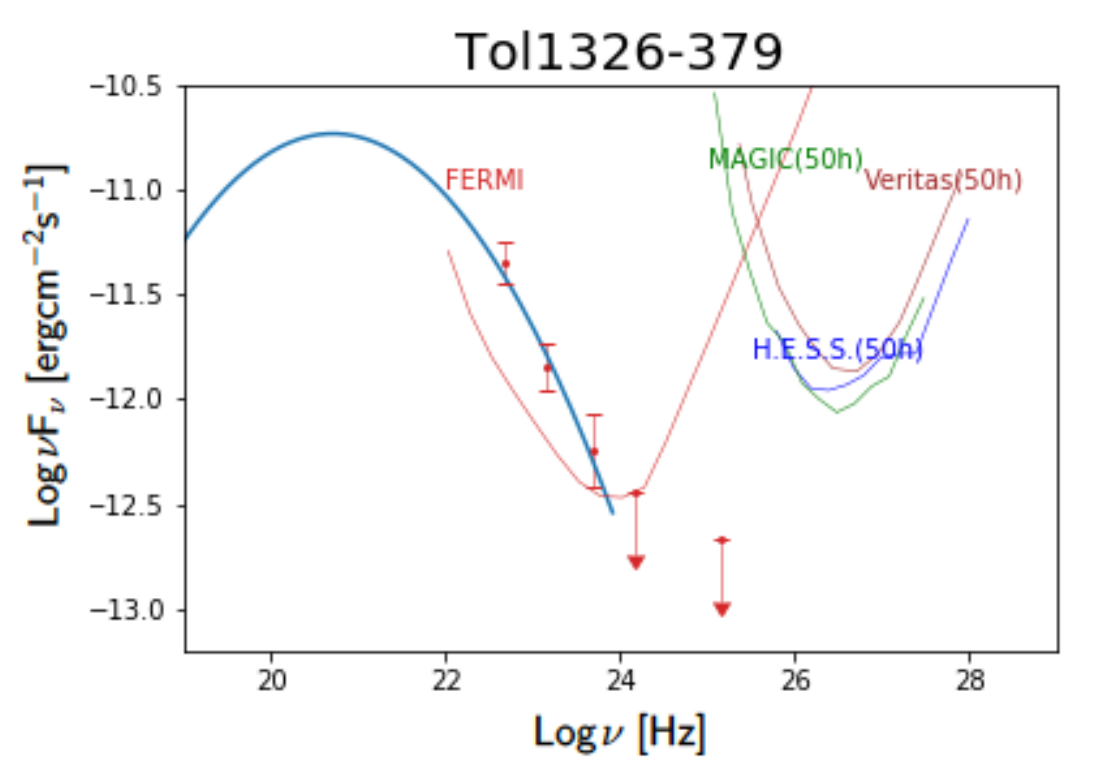}
\caption{SED of Tol~1326~$-$~379 in the~high energy band. The~red points are Fermi-LAT measurements (with the corresponding sensitivity curve) fitted by a~power-law component that reproduces the~inverse Compton bump. We plot the~expected differential sensitivity curves for~50~h of exposure of {\it MAGIC} (in~green)~\cite{sitarek15}, {\it VERITAS} (in~brown)~\cite{park15} and~{\it H.E.S.S} telescopes (in~blue)~\cite{holler15}.}
\label{tolsed}
\end{figure}

In a~TeV regime, we compare the~flux sensitivity curves of current Cherenkov telescopes ({\it MAGIC}~\cite{sitarek15}, {\it VERITAS}~\cite{park15}, {\it H.E.S.S.}~\cite{holler15}) with the~SEDs of the~two FR~0s cases (Figure~\ref{FR0SED}). The~Inverse Compton bump is modelled with a~power-law component~\cite{grandi16} that fits the~measured points in the~{\it Fermi} band; above~10$^{24}$~Hz the~source is undetected. The~predicted SED in the~VHE regime is well below the~sensitivity of all the~Cherenkov telescopes in~50~h of observation (see Figure~\ref{tolsed} for a~zoom of the~SED of Tol~1326~$-$~379 at~VHE). A major step further will be  possible thanks to the~forthcoming Cherenkov Telescope Array (CTA)~\cite{actis11}.  Thanks to the~improvement in sensitivity by a~factor 5--20 (depending~on the~energy range 10~GeV--300~TeV) with respect to current Cherenkov telescopes, CTA will enable population studies and~allow us to verify whether or not FR~0s are VHE emitters. Moreover, CTA will work as~an~observatory in synergy with other HE facilities, providing simultaneous data from a~few MeV up to hundreds of~TeV. This will guarantee the~opportunity to test different mechanisms of VHE photon production in RGs in general~\cite{angioni17,rieger18} and~FR~0s in particular, as~outlined in the~next section.

\section{Are FR~0s VHE Candidate Sources?}

The~lack of evident FR~0s in the~{\it Fermi/LAT} catalogues and~in the~TeV catalogues (e.g.,~TevCat and~TeGeV catalogues~\footnote{TevCat (\url{http://tevcat.uchicago.edu/}) and~TGeVCat (\url{https://www.ssdc.asi.it/tgevcat/}) are online catalogues for TeV~astronomy.}) is not surprising. However,~it~is worth noting that the~current Imaging Atmospheric Cherenkov Telescopes ({\it IACT}) operate in pointing mode, generally triggered by external alerts (e.g.,~the observation of an~optical, X-ray or $\gamma$-ray~flare). Consequently, their~TeV catalogues might include a~bias towards blazars.

The~question is therefore whether these sources may or may not be considered candidate~TeV emitters, and~which scenarios would imply their detection. In the~case of Tol~1326~$-$~379, a~detection $>$100~GeV would indicate the~presence of a~distinct radiative component, with respect to the~SSC~one, which should remain sub-dominant in the~0.1--10~GeV band.
Thus far, indications that such components could actually exist have been only found in the~SED of the~core of the~radio galaxy Centaurus A~\cite{aharonian09,sahakyan13}. Interestingly, its emission deviates  from the~simple extrapolation of the~$<$3~GeV spectrum observed by {\it Fermi}, displaying a~clear hardening at~$<$250~GeV (\cite{hess18} and~references~therein).  A~comprehensive discussion concerning the~processes that could explain the~origin of a~TeV component is presented in~\cite{rieger18}. Here, we review the~different models of VHE photon production in AGN in relation to the~FR~0~class.

The~conventional leptonic jet (one-zone) SSC models predicts that X-ray-to-$\gamma$-ray part of the~SED  originates from inverse Compton (IC) scattering of low energy photons by relativistic electrons in a~single emitting region within the~inner jet. The~scattered photons may have an~internal origin, produced by the~synchrotron mechanism (SSC,~\cite{maraschi92}), or external origin, i.e.,~seed photons coming from the~accretion disc, the~broad line region, or the~dusty torus (external Compton, EC, e.g.,~\cite{dermer95,ghisellini96}).   Still~in the~framework of leptonic models, a~jet with a~stratified kinematic structure opens to multi-zone emission models. In its simplest realisation, this could be a~two-zone model with a~inner fast spine ($\Gamma_1  \gg$ 1) nested in a~outer slower sheath ($\Gamma_2 < \Gamma_1$)~\cite{ghisellini05}. Depending on the~observer's line of sight, the~emission of one of the~two zones prevails, with the~spine dominating at~small angles (i.e.,~in~blazars) and~the~layer at~larger angles (in misaligned AGN). The~two emitting regions may radiatively interact, by providing each other additional target synchrotron photons to be Compton scattered. Because of the~velocity gradient between the~two emitting regions, the~synchrotron emission coming from one region is boosted in the~other, making this an~efficient way to produce HE emission.

Hadronic models, in which relativistic protons within the~jet are ultimately responsible for the~observed emission, have been also proposed (e.g.,~\cite{bottcher07,rani19b}). These relativistic protons can interact with the~soft synchrotron radiation and~produce electromagnetic cascades above several tens of~TeV and~be responsible for the~neutrino production. FR~0s have been recently proposed as~cosmic neutrino emitters because of their $\gamma$-ray emission~\cite{jacobsen15,tavecchio18}.

In the~BH proximity, the~BH gap model predicts that particles streaming along the~magnetic field lines in the~magnetosphere can be accelerated to very high energies due to the~formation of (partially) unscreened electric field regions (gaps)~\cite{ruderman75,gil03}. The~width of the~gap $h$ and~the~BH mass set the~maximum $\gamma$-ray luminosity that can be emitted with Compton peaks in the~range 10--100~TeV. In this gap model (stationary or time-dependent), very low accretion rates, or the~presence of a~radiatively-inefficient disc, represent a~necessary condition for the~detectability of magnetospheric VHE emission. This
requires the~accretion rate to be below a~critical value, typically $\sim$0.01 (e.g.,~\cite{yuan04}) and~in turn leads to a~constraint on the~average jet power. Assuming a~rapidly spinning BH, the~constraints on the~gap size and~accretion rate thus translate into a~characteristic upper limit on the~extractable VHE power of \vspace{2pt} $L_{gap}^{VHE} \propto 2 \times 10^{46} {\dot m_{E}} \,\ M_{BH} (h/r_{g})$ ~\cite{katsoulakos18} where ${\dot m_{E}}$ is the~accretion rate scaled at~the~rate at~the~Eddington limit and~r$_{g}$ is the~gravitational radius. Assuming that the~gap width in FR~0s is consistent with the~gravitational radius as~found from the~VHE variability of M~87 (e.g.,~\cite{albert08}) and~assuming similar accretion rates to M~87 as~supported by the~results for CoreG~\cite{balmaverde08} and~for FR~0s~\cite{torresi18a}, the~expected VHE luminosities for FR~0s from the~gap models are $10^{42\text{--}43}$~erg~s$^{-1}$. For~M87, the~VHE luminosity is~3--10~$\times$~10$^{40}$ erg s$^{-1}$ (e.g.,~\cite{aharonian03}), which~is three orders below the~expected VHE gap luminosity.  Therefore,~since~FR~0s and~M~87 have similar bolometric AGN power, we could expect~TeV luminosities for the~FR~0s to be an~order of magnitude lower because of their on average one-magnitude lower BH masses ($\sim$$10^{8} M_{\odot}$) than~M~87.

\begin{figure}[H]
\centering
\includegraphics[width=0.4\textwidth,angle=90]{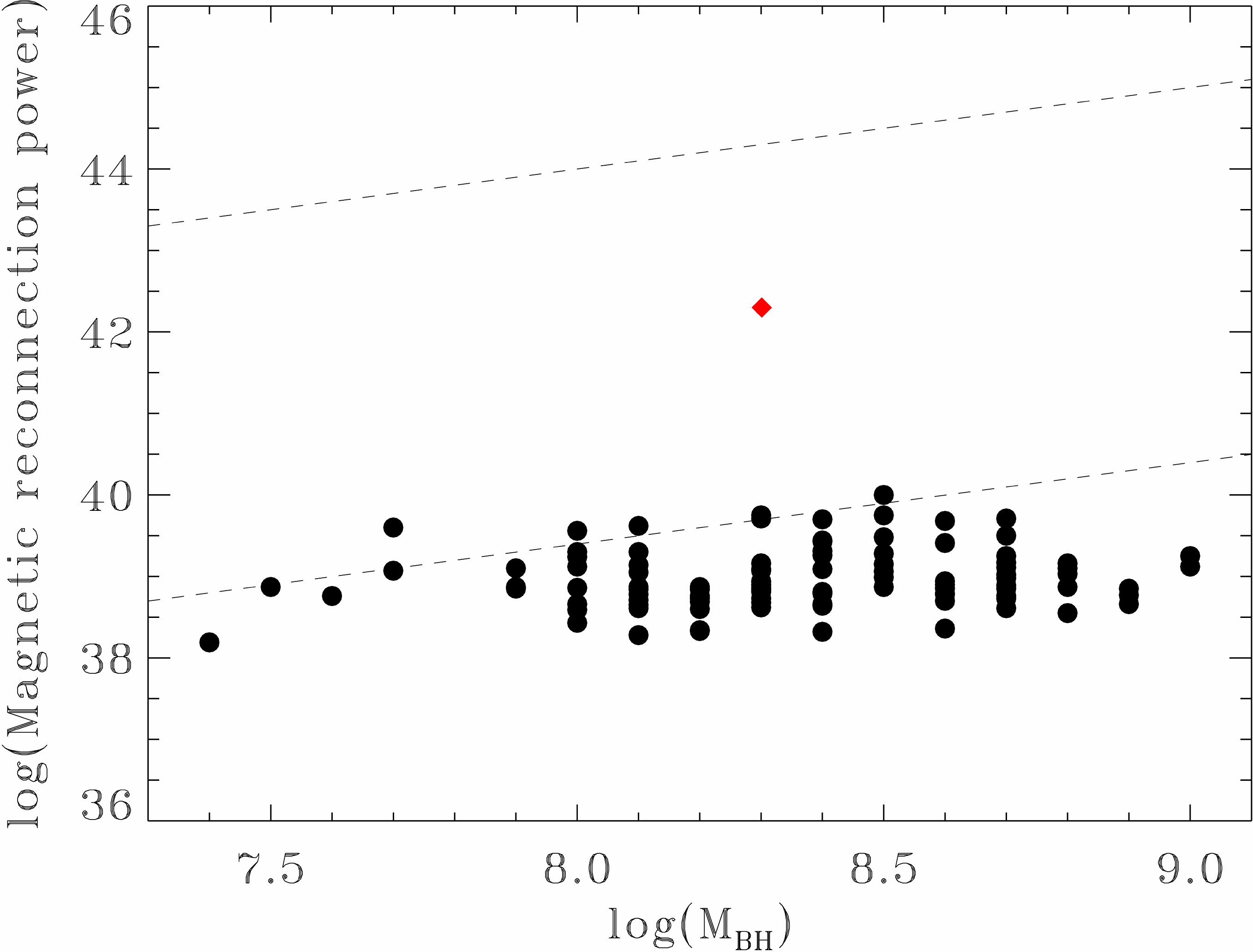}
\caption{Turbulent driven magnetic reconnection power (erg s$^{-1}$) against BH masses (M$_{\odot}$) compared to the~observed radio emission from FIRST catalogue for the~104 FR~0s from the~FR0{\sl{CAT}} (black points). For Tol~1326~$-$~379, we also plot the~high-energy ($>$1~GeV) luminosity (red diamond). The~region enclosed between the~two dashed lines represents the~maximum range of magnetic reconnection power predicted for RGs for supermassive BHs with different internal conditions (see~\cite{kadowaki15}).}
\label{magneticpower}
\end{figure}

Relativistic jets are generally expected to be initially magnetically supported and~models of collisionless magnetic reconnection are known to accelerate non-thermal electrons and~produce HE
photons by IC scattering (e.g.,~\cite{nalewajko11,cui12}) in the~jets. Magnetic reconnection energy, when released, may provide an~additional relativistic velocity component of the~ejected plasma relative to the~mean bulk flow of the~jet, in any direction. This Doppler-boosted effect may lead to VHE emission even in misaligned AGN, including FR~0s.
If the~released magnetic power is responsible for the~acceleration of the~radiating relativistic particles, then the~radio synchrotron radiation sets the~minimum value to this power. Figure~\ref{magneticpower} compares the~magnetic reconnection  power driven by  turbulence derived by~\cite{kadowaki15} with the observed core radio luminosities of FR~0s taken from the~FR0{\sl{CAT}}~\cite{baldi18}. We find that the~magnetic reconnection power extracted from reconnection of the~magnetic lines in the~inner coronal region around the~BHs of FR0s could be energetically sufficient to explain the~core synchrotron radio emission from them, as~also found valid  for low-luminosity AGN~\cite{degouveia10,kadowaki15}.  Figure~\ref{magneticpower} also includes the~$\gamma$-ray luminosity of Tol~1326-379, indicating that the~magnetic reconnection produces enough power to emit HE $\gamma$-ray radiation.

 The~scenario where $\gamma$-ray emission is produced by isotropically distributed relativistic electrons in the~kpc-scale jet~\cite{bednarek19} has been raised to reconcile with the~extended HE detection from the~jet lobes of CenA~\cite{abdo10b} and~with the~possibility that X-ray jets could emit at~VHE~\cite{hardcastle11,liu17}. Relativistic electrons in the~extended jets or lobes could Comptonise soft non-thermal X-ray radiation and~CMB photons, which are
beamed due to electron relativistic motion. The~mild relativistic jet speed at~a kpc scale expected for the~FR~0s and~their limited jet sizes disfavour but do not fully reject the~possibility of VHE emission originated on a large jet scale from the~FR~0 population.

On the~whole, the~VHE-emission models predict that FR~0s are~TeV emitters with luminosities comparable or within an~order of magnitude less than those observed from the~core of local FR~Is (i.e.,~M~87 and~CenA). Nevertheless,~the~low brightness and~further distance of the~FR~0 population than the~known~TeV-emitting misaligned RGs (at~least a~factor of~2 in~distance) hampers their detection at~VHE even with future sensitive instruments. All the~more, because it is unexpected, the~detection of even a~single FR~0 would open up interesting scenarios on the~jet physics (radiative~processes and~jet~accelerations) in these peculiar RGs.

\section{Summary, Conclusions and~Future Perspectives}

The~ongoing progress over the~last few decades with consecutive {\it Fermi} catalogues and~the~upcoming advent of high-sensitivity MeV-TeV facilities make the~HE astronomy accessible to a large part of the~community and~these observatories play a~unique scientific role in the~comprehension of the~jet physics in radio-loud AGN. In such a~framework, FR~0s, compact RGs with a substantial deficit of extended radio emission~\cite{baldi16}, are catching the~attention of the~community because of their large abundance with respect to the~other FR classes and~their potentiality of being HE emitters as~a~new~class.

 In this work, we reviewed the~multi-band properties of this class of RGs, from radio to X-rays. Then, based on the~unique {\it Fermi} detection of the~FR~0 population (i.e.,~Tol~1326-379~\cite{grandi16}) so far, we~explored the~possibility of~GeV--TeV emission of this abundant class of RGs, considering current (conventional and~more exotic) models for $\gamma$-ray emission in AGN. Indeed,~the~{\it Fermi} detection of an~FR~0 has opened up several scenarios to interpret its multi-wavelength SED. Whereas the~similar broadband spectral nuclear properties of FR~0s and~FR~Is suggest common jet physics in the~proximity to the~BH, the~indisputable differences of the~jet structures at~large scales between the~two classes might imply intrinsic variations in the~particle accelerations and~in HE emitting processes. A weak/short fast spine and~the~scarcity of sites of particle accelerations, limited to several pc along the~jet, reconcile with the~global multiband properties of FR~0s, the~confined jet structures and~the~kpc-scale jet symmetry. In~a~time-dependent scenario, a~nuclear recurrent activity could lead to a~discontinuous plasma injection in the~jet and~its possible deceleration. In the~framework of HE emitting sites, the~core region appears to be most plausible scene where $\gamma$-ray radiation can be produced because of the~dominant role of the~fast spine in the~jet, which can relativistically accelerate particles.

Several mechanisms can be proposed to address the~particle acceleration in the~inner part of the~FR0 jets and~their $\gamma$-ray production. Relativistic shocks between jet layers moving at~different speeds are responsible for accelerating particles along the~shock front and~could produce significant $\gamma$-ray emission by IC. A magnetospheric gap model followed by a~radiatively inefficient accretion flow to avoid $\gamma\gamma$-pair production seems to reconcile with the~expected HE luminosities of FR~0s. Magnetic reconnection, where magnetic energy is transferred to accelerating particles, predicts non-thermal synchrotron powers that are sufficient to justify the~observed radio core luminosities of FR~0s.

Detection of new FR~0s at~HE and~VHE bands in the~next future or stringent upper limits could help to set the~jet parameter space of this peculiar class of RGs and~constrain models on particle acceleration and~HE emission processes. In this framework, to test whether FR~0s could be detected at~HE, we compare the~broadband SEDs of two cases of FR~0s with the~sensitivity curves of MeV--GeV satellites and~TeV Cherenkov telescopes. At MeV energies, the~possible detection of the~FR~0s will cover the~gap between the~X-ray and~the~VHE emission, crucial for having a~comprehensive view of the~jet properties. Below a~few~GeV, the~detection of new FR~0s appears to be feasible with current and~future $\gamma$-ray facilities, and~more sources could possibly hide among the~unknown blazar candidates in the~{\it Fermi} catalogue. At~VHE, the~steep~GeV spectrum and~low brightness of Tol~1326~$-$~379 make a~detection unlikely. However,~our~knowledge of this class at~high energies is still very limited, hence it will be important to investigate the~FR 0s with more sensitive observatories, such as~CTA in the near~future.

The~FR~0 population, which outnumbers the~FR~Is, is expected to contribute up to $\sim$18\% to the~extra-galactic $\gamma$-ray background~\cite{stecker19}. Recently,~TeV-emitting candidates associated with low-power compact radio sources hosted in red ETGs have been selected by \cite{balmaverdeterex}. The~authors predict that a~few hundred of these candidates will be detectable by new generations of Cherenkov telescopes in~50~h in the~entire extra-galactic sky in the~local Universe. We~speculate that these~TeV emitters could be the~parent aligned population of the~FR~0s .

 The~optimal already-tested synergy between current radio, X-ray and~$\gamma$-ray telescopes in the~last decade and~the~upcoming generations of high-sensitivity and~high-resolution facilities in such bands (i.e, {\it SKA}, {\it LOFAR}, {\it ngVLA}, {\it eROSITA}, {\it Athena}, {\it Lynx}, {\it CTA}) will offer unique insights into jet physics and~particle acceleration mechanisms in the~vicinity of the~BH. These new  observatories will probe the~very low and~high energy bands of the~SEDs of the~RGs which populate the~low luminosity tail of the~local radio AGN luminosity~\cite{best12}. In~this very low-power regime, an~enormous population of compact radio sources, which  generally includes all nearby giant elliptical galaxies~\cite{nagar00,falcke00,filho02,nyland16,baldi18b} consistent with an~FR~0 classification, could accelerate jets and~potentially be $\gamma$-ray emitters. CTA,~which will unearth such a~population of low-power HE sources, will provide important progress in our understanding of the~AGN phenomena, by providing robust constraints on the~BH-jet coupling at~a low-luminosity regime.


\vspace{6pt}



\authorcontributions{ R.D. B. has led  several reference works on FR~0s, developing Sec. 1, 2.2, 3, 5, and 6, and coordinated this review. E.T. has led the work on the results presented in Sec. 2.1 and contributed to Sec. 4, G.M. and B.B. has led the discussion in Sec. 4. All the authors contributed to the discussion and writing of this work.} 

\funding{R.D.B. acknowledges the~support of STFC under grant ST/R000638/1 and~the~Italian grant (Progetto di Ateneo/CSP~2016). E.T acknowledges the financial contribution from the agreement ASI-INAF n.2017-14-H.0.}

\acknowledgments{We thank Alessandro Capetti and Paola Grandi for the~discussions that have contributed
to the~ideas that we expose in this review, and~for sharing the~multi-band data for the~SED of Tol~1326-379. We thank the~referees for providing useful comments to the~original manuscript and X.P. Cheng for giving us permission to add his figure into this publication.
Part of this work is based on archival data, software or online services provided by the Space Science Data Center - ASI.
This research has made used of TOPCAT\footnote{http://www.star.bris.ac.uk/~mbt/topcat/} \cite{taylor05} for the preparation and manipulation of the tabular data.
}

\conflictsofinterest{The~authors declare no conflict of interest. The~founders had no role in the~design of the~study; in the~collection, analyses, or interpretation of data; in the~writing of the~manuscript, or in the~decision to publish the~results.}

\appendixtitles{no} 
\appendixsections{multiple} 



\reftitle{References}





\end{document}